\documentclass[pra,preprint,preprintnumbers,amsmath,amssymb,floatfix]{revtex4}

\usepackage{graphicx}
\usepackage{dcolumn}
\usepackage{bm}
\usepackage{color}

\newcommand{\dyad}[1]
    {\overset{\lower0.5em\hbox{$\smash{\scriptscriptstyle\leftrightarrow}$}} {#1}}

\begin{document}

\title{Self alignment and instability of waveguides induced by optical forces}

\author{Amit Mizrahi}
\email{amitmiz@ece.ucsd.edu}
\author{Kazuhiro Ikeda} \author{Fabio Bonomelli} \author{Vitaliy Lomakin} \author{Yeshaiahu Fainman}
\affiliation{Department of Electrical and Computer Engineering,
University of California, San Diego\\ 9500 Gilman Drive, La Jolla, California 92093-0407, USA}

\begin{abstract}
We introduce a new fundamental property of waveguides induced by the forces of the guided light, namely, the ability to self align or be in instability.
A nanoscale waveguide broken by an offset and a gap may tend to self align to form a continuous waveguide.
Conversely, depending on the geometry and light polarization, the two parts of the waveguide may be deflected away
from each other, thus being in an unstable state.
These effects are unique as they rely on the presence of both the guided mode and the scattered light.
Strong self alignment forces may be facilitated by near field interaction with polarization surface charges.
\end{abstract}

\maketitle

\section{Introduction}

Laser light has significant mechanical effects on microscopic objects, as was initially pointed
out about four decades ago~\cite{Ashkin:1970(Acceleration_and_trapping)}.
In addition to the vast work on trapping and manipulation of small particles~\cite{grier2003rom},
much effort has been directed at cavity based optomechanical devices, where
optical forces may be considerably enhanced~\cite{notomi2006owa,kippenberg2008cob,tomes:113601}.
A less explored option, however, is optical forces on waveguides, which become observable
at the microscopic scale~\cite{Painter:Jul2007(Actuation),she2008opf}.

Research of optical forces on waveguides is motivated by
the growing capabilities of nanofabrication that enable new
possibilities of nanoscale light manipulation~\cite{Lipson:2004(Exper_Guiding_in_void),levy2005idm,Tan:2008(chip-scale)}.
The theory of guided light is thus being extended to include the laws of the mechanical effects of light
on the guiding structure itself~\cite{Povinelli:2004(Slow-light),Povinelli:2005(Evanescent_bonding),Mizrahi:2005(Mirror_Manipulation),Mizrahi:2006(Accelerator_forces),Mizrahi:2007(Two_slab_spring),Mizrahi:Aug2008(rotating_modes)}.
These new physical mechanisms relate the properties of the guided modes to the forces created by them.
For instance, light guided between two waveguides or mirrors creates a repulsive force for an antisymmetric transverse field
and an attractive force for a symmetric transverse field~\cite{Povinelli:2005(Evanescent_bonding),Mizrahi:2005(Mirror_Manipulation)}.
Moreover, a superposition of a symmetric and antisymmetric modes may hold the waveguide in a stable
equilibrium~\cite{Mizrahi:2005(Mirror_Manipulation),Mizrahi:2007(Two_slab_spring),rakich2007tca}.
Such phenomena can be experimentally observed in nanomechanical devices fabricated on a chip, as was
recently demonstrated with a suspended silicon waveguide~\cite{Li:2008(nature_harnessing)}.

In this study, we introduce a new fundamental property of waveguides, namely the ability
of a waveguide to self align by light forces when it is perturbed by a small offset misalignment.
We show that the size of the waveguide and the type of eigenmode determine whether the two misaligned parts
will tend to self align or deflect away from each other.
This phenomenon is unique as it relies on both the \emph{guided} waveguide eigenmode and the
\emph{scattered} radiation from the perturbation.
A strong self alignment force is created due to polarization surface charges that dominate the optical force with a near field interaction.
We further investigate the exerted forces when a gap is introduced between the two parts of the waveguide.

\begin{figure}[]
\centerline{\rotatebox{0} {\scalebox{1}{\includegraphics[width=8.6cm]{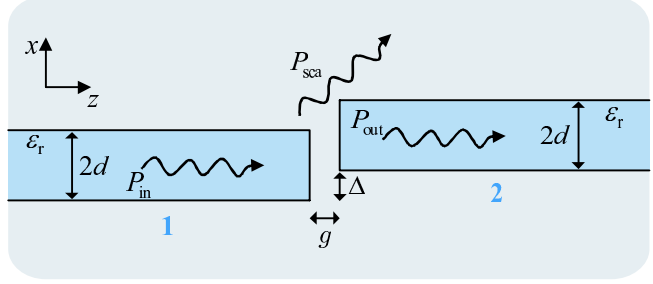}}}} \caption{\label{fig:wav_coup_config}
A slab waveguide broken by an offset and a gap.}
\end{figure}

The geometry under consideration is shown in Fig.~\ref{fig:wav_coup_config}.
A single mode slab waveguide of half-width $d$ and permittivity $\varepsilon_\mathrm{r}$ is broken by an offset in the $x$~axis, $\Delta$, and a
gap in the $z$~axis $g$.
An eigenmode is incident from the left (input waveguide) carrying power $P_\mathrm{in}$, most of which is transmitted
to the output waveguide ($P_\mathrm{out}$), while the remainder is either scattered ($P_\mathrm{sca}$) or reflected back into the input waveguide.
No variations of the geometry are assumed along the $y$ axis, and therefore all quantities given are per unit length.

\section{A waveguide perturbed by an offset}

First, we examine the case of no gap, $g=0$, and a small offset $\Delta$. For the calculation
of the fields we take an approach of mode matching approximation, similar to that described by
Marcuse~\cite{Marcuse:1970(radiation_losses),Marcuse:1972Book(Light_Transmission)}.
We begin by considering a transverse electric (TE) incident mode,
for which the nonzero field components are $E_y$, $H_x$, and $H_z$.
The even guided mode incident from the left is given inside
the dielectric slab by $E_\mathrm{i}=A_\mathrm{TE}\cos(k_xx)\exp(-j\beta_\mathrm{g} z)$,
where $A_\mathrm{TE}=\sqrt{2\omega\mu_0P_\mathrm{in}/\beta_\mathrm{g}(d+\gamma^{-1})}$,
$k_x$ is the transverse wavenumber in the dielectric,
$\beta_\mathrm{g}$ is the longitudinal wavenumber,
$\gamma$ is the transverse decay constant outside the slab,
and the time dependence is of the form $e^{j\omega t}$.
The transverse electric field in each region may be represented as a sum of the guided modes
and the continuous spectrum of the radiation modes, corresponding to the scattered light.
Explicitly, assuming the interface plane is at $z=0$, then for $z<0$ (input waveguide) this field is given by
\begin{equation} \label{eq:TE_Left}
E_1=E_\mathrm{i} + a_\mathrm{r}E_\mathrm{r} +
\int_0^\infty\!\! \mathrm{d}\rho \, q_{\mathrm{e}1}(\rho)E_{\mathrm{e}1}+
\int_0^\infty\!\! \mathrm{d}\rho \, q_{\mathrm{o}1}(\rho)E_{\mathrm{o}1} \,,
\end{equation}
whereas for $z>0$ (output waveguide) it reads
\begin{equation} \label{eq:TE_Right}
E_2=a_\mathrm{t}E_\mathrm{t} +
\int_0^\infty\!\! \mathrm{d}\rho \, q_{\mathrm{e}2}(\rho)E_{\mathrm{e}2}+
\int_0^\infty\!\! \mathrm{d}\rho \, q_{\mathrm{o}2}(\rho)E_{\mathrm{o}2} \,.
\end{equation}
In the above two equations, $a_\mathrm{r}$ and $a_\mathrm{t}$ are the reflection and transmission coefficients
of the guided mode, respectively; $E_\mathrm{r}$ and $E_\mathrm{t}$ are the reflected and transmitted guided
modes, respectively; $q_{\mathrm{e}1,2}$ and $q_{\mathrm{o}1,2}$ are the amplitudes of the
even and odd radiation modes, respectively; $\rho$ is the transverse wavenumber of the radiation
modes outside the slabs; $E_{\mathrm{e}1,2}$ and $E_{\mathrm{o}1,2}$ are the even and odd radiation
modes, respectively. The amplitudes of the guided and the radiation modes may be approximated
analytically by expressions containing the overlap integral between the respective mode and the incident guided mode.

Generally, force densities on dielectrics may be viewed as resulting from two
processes~\cite{Mizrahi:2005(Mirror_Manipulation),Schwinger:1998Book(Classical_Electrodynamics),Manusripur:2004(Radiation_pressure)}: (1) the interaction of effective polarization
\emph{volume} current densities with the magnetic field, and (2) the interaction of polarization \emph{surface} charge densities with the electric field.
In the TE case,
no component of the electric field is perpendicular to the boundaries of the dielectrics, and
therefore no polarization surface charge densities are formed.
Thus the time-averaged volume force density in the $x$ direction is given by
$\frac{1}{2}\mathrm{Re}\left(j\omega P_y\times\mu_0{H_z}^*\right)$, where $P_y=\varepsilon_0(\varepsilon_\mathrm{r}-1)E_y$ is
the polarization density.
By integration over the volume, we obtain the total transverse force on the output waveguide
in terms of integration over only the top and bottom surfaces
\begin{equation} \label{eq:Force_TE}
F_{x2}=\frac{1}{4}\varepsilon_0(\varepsilon_\mathrm{r}-1)
\int_0^\infty\!\!\mathrm{d}z\left(
\left.|E_2|^2\right|_{x=d+\Delta}-
\left.|E_2|^2\right|_{x=-d+\Delta}\right) \,.
\end{equation}
When there is no offset ($\Delta=0$), the field is symmetric around $x=0$ and the guided mode is not disturbed by
a discontinuity, and thus no force acts on each of the waveguide parts.
Once an offset is introduced,  symmetry is broken, and scattering occurs from the discontinuity.
The question then arises whether the two parts of the waveguide will tend to deflect away from each other, or self-align
to form back a continuous waveguide while maximizing the output power.

For the evaluation of the force on the output waveguide, the field expression of Eq.~\eqref{eq:TE_Right} is substituted
into Eq.~\eqref{eq:Force_TE}.
At this point, we are interested in small offsets, so that only terms up to the first order of $\Delta$ are kept.
Noting that cross-products of the even radiation modes with the odd radiation modes are of order larger than
$\Delta$, and using the symmetry properties of the modes, the expression for the force reads
\begin{equation} \label{eq:Fx2_TE}
F_{x2}\simeq\varepsilon_0(\varepsilon_\mathrm{r}-1)\mathrm{Re}
\int_0^\infty\!\!\mathrm{d}z\,
{E_\mathrm{t}}^*\int_0^\infty\!\! \mathrm{d}\rho \, {q_{\mathrm{o}2}(\rho)}{E_{\mathrm{o}2}}  \,,
\end{equation}
where the integration is performed at $x=d+\Delta$.
Hence, it is evident that the odd radiation modes created by the scattering are responsible for the transverse force on the
waveguide.
A direct measure of the waveguide's tendency to move either way is the derivative of the force with respect to $\Delta$,
$\mathrm{d}F_{x2}/\mathrm{d}\Delta(\Delta=0)$ denoted by ${F_{x2}}'$.
Bearing in mind that $F_{x2}(\Delta=0)=0$, the force may be approximated by ${F_{x2}}'\Delta$.
The only term in the above equation that depends on $\Delta$ is
$q_{\mathrm{o}2}$, and an analytic expression for $\mathrm{d}q_{\mathrm{o}2}/\mathrm{d}\Delta$ may be obtained.
The integration over $z$ is then performed analytically, and the closed form expression for the derivative reads
\begin{equation} \label{eq:dF_TE}
{F_{x2}}'=
\varepsilon_0(\varepsilon_\mathrm{r}-1)^2k_0^2
A_\mathrm{TE}^2\cos^2(k_xd)/\pi
\int_0^\infty\!\! \mathrm{d}\rho \, \frac{1}{1+\frac{\sigma^2}{\rho^2}\cot^2(\sigma d)}
\frac{1}{\gamma^2+\rho^2} \mathrm{Re}
\frac{j(\beta_\mathrm{r}+\beta_\mathrm{g})}{\beta_\mathrm{r}(\beta_\mathrm{r}-\beta_\mathrm{g})}\,,
\end{equation}
where $k_0=\omega/c$.
While the spectrum of radiation modes contains both propagating modes having real $\beta_\mathrm{r}$
and evanescent modes having imaginary $\beta_\mathrm{r}$, the above expression shows that it is only
the evanescent radiation modes that contribute to the generation of this force, i.e., the integrand is nonzero
only for $\rho>k_0$.

\begin{figure}[]
\centerline{\rotatebox{0} {\scalebox{1}{\includegraphics[width=6.6cm]{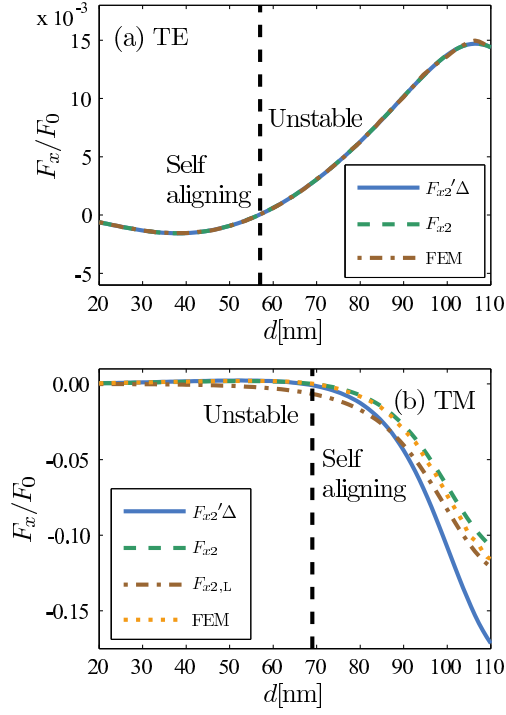}}}} \caption{\label{fig:small_offset_Fx}
Transverse force as a function of $d$ for an offset $\Delta$ of 2\% of $d$. The permittivity of the waveguides
is $\varepsilon_\mathrm{r}=3.48^2$, and the wavelength is $\lambda=1.55$~$\mu$m. (a) TE incident mode. (b) TM incident mode.}
\end{figure}

Both the force $F_{x2}$ [Eq.~\eqref{eq:Fx2_TE}] and the quantity ${F_{x2}}'\Delta$ [Eq.~\eqref{eq:dF_TE}] are plotted in
Fig.~\ref{fig:small_offset_Fx}(a) as a function of $d$, for an offset of 2\% of $d$.
The range of $d$ shown is 20~nm to 110~nm, where the slab is single mode in each polarization.
The wavelength is taken to be $\lambda=1.55$~$\mu$m and the permittivity is $3.48^2$, corresponding to silicon at that wavelength.
The forces are normalized by $F_0\equiv P_\mathrm{in}/c$, which is the momentum per unit time carried by a plane wave.
These results are compared with a Finite Element Method (FEM) simulation, and as seen the three curves are virtually identical.
In addition to the integration of the force on the polarization densities, we have also integrated over the Maxwell stress tensor~\cite{Schwinger:1998Book(Classical_Electrodynamics)}
to obtain the force,
and found excellent agreement between the two methods, which are mathematically equivalent for exact solutions of Maxwell's equations.

\begin{figure}[]
\centerline{\rotatebox{0} {\scalebox{1}{\includegraphics[width=8.6cm]{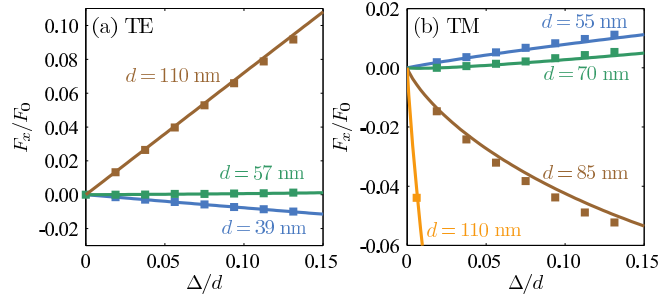}}}} \caption{\label{fig:F_DEL}
Transverse force $F_{x2}$ as a function of the offset $\Delta$ for different values of waveguide half-width
$d$. The solid line corresponds to the analytic analysis and the square markers indicate FEM simulation results.
(a) TE incident mode. (b) TM incident mode.}
\end{figure}

The plot of Fig.~\ref{fig:small_offset_Fx}(a)
reveals two regimes; the first is a self-alignment regime up to a slab half-width of about 57~nm, for which ${F_{x2}}'<0$
corresponding to a restoring force.
The second is an instability regime in which a small offset results in a deflection force.
Both regimes exhibit an optimal slab width for which the force is strongest.
Although we show here only the transverse force on the output waveguide, when the offset is small, the scattered power is negligible and
by virtue of momentum conservation, the force on the input waveguide is of the same magnitude and opposite in sign to that on the output waveguide.
In fact, we have shown analytically that $F_{x1}'$=-$F_{x2}'$.
To better illustrate the different regimes, we depict in Fig.~\ref{fig:F_DEL}(a) the transverse force $F_{x2}$ as function of
$\Delta$ for $d=39$~nm where a negative restoring force is seen, $d=57$~nm where the force derivative at $\Delta=0$ vanishes
at the transition between the two regimes, and for $d=110$~nm where instability in the form of a deflecting force is observed.

When the incident mode is transverse magnetic (TM), the situation is considerably more involved.
The field components for the TM mode are $E_x$, $E_z$, and $H_y$, and the incident
magnetic field is given by $H_\mathrm{i}=A_\mathrm{TM}\cos(k_xx)\exp(-j\beta_\mathrm{g}z)$, where
\begin{equation*}
A_\mathrm{TM}=\sqrt{2\omega\varepsilon_0\varepsilon_\mathrm{r}P_\mathrm{in}/\beta_\mathrm{g}[d+\varepsilon_\mathrm{r}\gamma^{-1}(k_x^2+\gamma^2)/(k_x^2+\varepsilon_\mathrm{r}^2\gamma^2)]}.
\end{equation*}
The fields in each region are described by
Eqs.~\eqref{eq:TE_Left}~and~\eqref{eq:TE_Right} with $E$ replaced by $H$.
The force mechanism in the TM case differs substantially than that of the TE case, as for the TM
the electric field has normal components to discontinuities in the dielectric, and therefore
polarization surface charge densities are formed. These surface charge densities interact
with $E_x$ to give rise to surface force densities.
Considering again the output waveguide, the first contribution is from the polarization surface charge densities
created by $E_z$ at the $z=0$ interface. The force density of this term is given by
$\frac{1}{2}\mathrm{Re}[-\varepsilon_0(\varepsilon_\mathrm{r}-1)E_z{E_x}^*]$, where $E_z$ is the field just inside
the slab at $z=0$.
Calculation of this force on the output waveguide while keeping only terms up to the first order of $\Delta$ yields
\begin{multline} \label{eq:FxL_TM}
F_{x2,\mathrm{L}}\simeq
\frac{A_\mathrm{TM}}{2}\frac{\varepsilon_0(\varepsilon_\mathrm{r}-1)}{(\omega\varepsilon_0\varepsilon_\mathrm{r})^2}
\mathrm{Re}\,j\!
\int_0^\infty\!\! \mathrm{d}\rho \, q_{\mathrm{o}2}(\rho)\times\\
\left\{\frac{(k_x\beta_\mathrm{r}+\sigma\beta_\mathrm{g})\sin[(k_x-\sigma)d]}{(k_x-\sigma)}+
\frac{(k_x\beta_\mathrm{r}-\sigma\beta_\mathrm{g})\sin[(k_x+\sigma)d]}{(k_x+\sigma)}\right\}  \,,
\end{multline}
where an analytic expression for $q_{\mathrm{o}2}$ may be obtained.
A second contribution is from the polarization surface charge densities formed on the top ($x=d+\Delta$) and
bottom ($x=-d+\Delta$) parts of the waveguide,
given by the plus and minus of $\frac{1}{4}\varepsilon_0(\varepsilon_\mathrm{r}^2-1)|E_x|^2$, respectively.
In both cases $E_x$ is the field just inside the slab.
Summing up the two contributions and integrating over $z$ results in the expression
\begin{equation} \label{eq:FxUD TM}
F_{x2,\mathrm{UD}}\simeq
A_\mathrm{TM}\beta_\mathrm{g} \frac{\varepsilon_0(\varepsilon_\mathrm{r}^2-1)}{(\omega\varepsilon_0\varepsilon_\mathrm{r})^2} \cos(k_xd)\,
\mathrm{Re}
\int_0^\infty\!\! \mathrm{d}\rho \, q_{\mathrm{o}2}(\rho)\times
\beta_\mathrm{r}\sin(\sigma d)/\left[j(\beta_\mathrm{r}-\beta_\mathrm{g})\right] \,.
\end{equation}
The third contribution comes from the volume force density given by
$\frac{1}{2}\mathrm{Re}[-\varepsilon_0(\varepsilon_\mathrm{r}-1)j\omega E_z\mu_0{H_y}^*]$, and integration
over $x$ and $z$ gives
\begin{equation} \label{eq:FxV TM}
F_{x2,\mathrm{V}}\simeq
-A_\mathrm{TM}\frac{\mu_0(\varepsilon_\mathrm{r}-1)}{\varepsilon_\mathrm{r}} \cos(k_xd)\,
\mathrm{Re}
\int_0^\infty\!\! \mathrm{d}\rho \, q_{\mathrm{o}2}(\rho)\times
\sin(\sigma d)/\left[j(\beta_\mathrm{r}-\beta_\mathrm{g})\right] \,.
\end{equation}
The total force is the sum of all three contributions, and its derivative $F_{x2}'$ is obtained by analytically differentiating $q_{o2}$.
Similarly to the TE mode, only the evanescent part of the odd radiation mode spectrum participates in the generation
of the force, and the relation $F_{x1}'=-F_{x2}'$ holds as well.

The quantities ${F_{x2}}^\prime\Delta$, $F_{x2}$, as well as the force calculated by FEM, are shown in Fig.~\ref{fig:small_offset_Fx}(b)
as a function of $d$ for an offset $\Delta$ of 2\%.
Contrary to the TE case, here instability occurs for small values of $d$, and above about 70~nm there is self-alignment.
The restoring force increases monotonically, so that at the maximum value in the shown range of $d$, it is about
two orders of magnitude stronger than the peak value of the TE restoring force.
This dramatic difference is due to the presence of electric field components that are perpendicular to the dielectric
boundaries.
Specifically, at the left boundary of the output waveguide ($z=0$ plane), $E_z$ induces a polarization surface charge
density, while $E_x$ gives it a transverse kick.
The result, $F_{x2,\mathrm{L}}$ given by Eq.~\eqref{eq:FxL_TM}, is plotted in Fig.~\ref{fig:small_offset_Fx}(b).
This force is negative for the entire range of $d$ and is seen to comprise almost all of the total force
in the self alignment regime.
Qualitatively, it may be associated with a dipole induced by the guided mode which has $E_z\propto\sin(k_xd)$,
and the strength of the dipole per power increases with $d$ as the mode confinement increases.
At the $z=0$ interface, the dipole is roughly inverted, and consequently the two parts attract each other, in both the
transverse and longitudinal directions.
Moreover, this force grows rapidly as a function of $\Delta$ and is therefore responsible for the derivative approximation
being less accurate than for the TE case.
This is seen in Fig.~\ref{fig:F_DEL}(b) where $F_{x2}$ is plotted for four different values of $d$: $d=55$~nm where instability is
observed, $d=55$~nm about where the derivative vanishes at $\Delta=0$,
$d=85$~nm which exhibits self alignment, and $d=110$~nm, where there is strong self alignment which
is further discussed below.

\section{A waveguide broken by an offset and a gap}
\begin{figure}[]
\centerline{\rotatebox{0} {\scalebox{1}{\includegraphics[width=8.6cm]{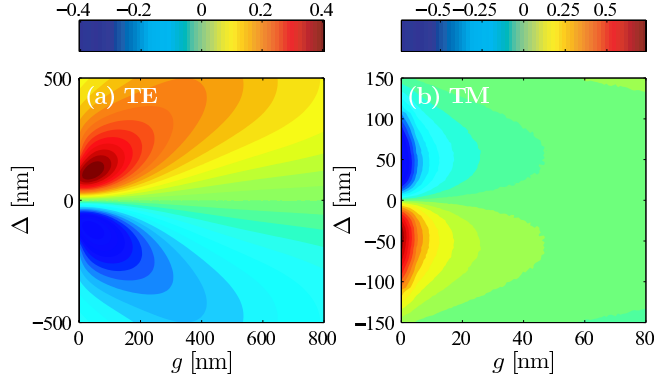}}}} \caption{\label{fig:cont_g_DEL}
Contours of $F_{x2}/F_0$ as a function of $g$ and $\Delta$ for $d=110$~nm and $\varepsilon_\mathrm{r}=3.48^2$.
(a) TE incident mode. (b) TM incident mode.}
\end{figure}

So far we have established the fundamental tendency of a waveguide to self-align or be in instability by considering
zero gap and a perturbation in a continuous waveguide in the form of an offset.
We next extend the discussion by introducing a longitudinal gap $g\neq0$.
Figure~\ref{fig:cont_g_DEL}(a) and Fig.~\ref{fig:cont_g_DEL}(b) show contours of $F_{x2}$ in the
$g$--$\Delta$ plane obtained by FEM simulations, for TE and TM incident modes, respectively.
The slab half-width is assumed to be $d=110$~nm, where
according to Fig.~\ref{fig:small_offset_Fx}, the TE mode places the system in
instability, whereas the TM mode causes self alignment.
In both cases the $g=0$ behavior extends to larger values of $g$, but the decay of the TE repulsive force with
the offset and gap is much slower than that of the attractive TM force, as seen by the scales of $g$ and $\Delta$ of the
two frames of Fig.~\ref{fig:small_offset_Fx}.
The maximal TE force is about $0.4F_0$ and it is obtained for about $\Delta\simeq130$~nm and $g\simeq50$~nm.
A waveguide cantilever at $g=0$ is, in fact, in a bistable state where a small offset may result in a deflection
force that would eventually be balanced by the mechanical force.
The attractive TM force is obtained for $g=0$ and $\Delta\simeq45$~nm, and it is about $0.7F_0$.
We further found that a strong longitudinal force is pulling the two waveguides towards each other at
a force of about $2F_0$ for $g\simeq20$~nm.
For large enough values of $g$ the TM force becomes repulsive, corresponding to radiation pressure.

The self alignment and the instability may be tested experimentally by fabricating
on a chip two waveguide cantilevers with an offset and a gap.
For instance, for $P_\mathrm{in}=30$~mW, $F_0=100$~pN, and at $g=20$~nm we obtain
$F_{x2}\simeq0.1F_0$ according to Fig.~\ref{fig:small_offset_Fx}(b), which is about 10pN.
This is of the order of magnitude of force that
was shown to actuate a silicon cantilever~\cite{Li:2008(nature_harnessing)}.
The deflection of the waveguide in such a system may be viewed by the nonlinear input/output behavior,
as the output power increases when the cantilevers tend to self align.
Moreover, the mechanical effect is doubled by the fact that a similar force opposite in sign is exerted
on both cantilevers.
The two cantilevers may also be vibrated at their mechanical resonance by modulating the incident power, resulting
in a system that may be suitable for applications such as sensing.

\section{Conclusion}

In conclusion, we demonstrated a novel effect of light forces in the form of
self alignment or instability of a waveguide broken by an offset and a gap.
The waveguide size and mode polarization determine which of the two regimes the waveguide is in.
Closed form expressions for the transverse forces were given for the case of a small offset and no gap.
The forces described here are unique as they are due to the presence of both the guided mode and the scattered light from the discontinuity.
Strong self alignment for a TM mode is caused by near field interaction of the polarization
surface charges created by the longitudinal electric field.
We are currently looking into the possibilities of an experimental realization that will demonstrate the 
effects discussed here.

\section*{ACKNOWLEDGMENTS}
This work was supported by the Defense Advanced Research Projects Agency, the National Science Foundation, the NSF CIAN ERC,
the U.S. Air Force Office of Scientific Research,
the U.S. Army Research Office, and the Technion Viterbi Fellowship.

\end{document}